\documentclass[amsmath,amssymb,showpacs,showkeywords,twocolumn]{revtex4}
\usepackage[dvips]{graphicx,color}
\usepackage{times}
\usepackage{mathrsfs}
\usepackage{amsmath}
\usepackage{graphicx}
\usepackage{epsfig}
\usepackage{dcolumn}
\usepackage{bm}
\usepackage{multirow}
\usepackage{xcolor}
\begin{document}
\title{Tunability of Andreev levels in a spin-active Ising Superconductor/Half Metal Josephson junction}
\author{Saumen Acharjee\footnote{saumenacharjee@dibru.ac.in}, Arindam Boruah\footnote{arindamboruah@dibru.ac.in}, Nimisha Dutta\footnote{nimishadutta@dibru.ac.in} and Reeta Devi\footnote{reetadevi@dibru.ac.in}}
\affiliation{Department of Physics, Dibrugarh University, Dibrugarh 786 004, 
Assam, India}

\begin{abstract}
We study the Andreev levels, supercurrent and tunnelling conductance in a clean Ising superconductor (ISC)/half metal (HM)/Ising superconductor (ISC) Josephson junction with spin-active interfaces using Bogoliubov-de Gennes equations. We theoretically demonstrate the effect of spin mixing, spin flipping processes and spin-orbit coupling (SOC) of the ISC on Andreev Bound States (ABS) spectra, current phase relation (CPR) and tunnelling conductance in transparent and opaque barrier limit. We witness an additional splitting of the Andreev levels due to SOC of the ISC and $0 - \pi$ transition for different barrier magnetic moments. Also, different $\phi$ - junctions can be achieved by tuning the strength of the barrier magnetic moment and spin mismatch angle. Moreover, a possible $0 - \pi$ transition can also be achieved for SOC stronger than the chemical potential of the ISC using suitable control parameters. The interplay of spin mixing and spin flipping processes with SOC can also host Majorana modes in the proposed system. The tunnelling conductance is found to be dependent on the spin mismatch angle. Also, a finite sub-gap conductance is observed, which indicates different probabilities of Andreev reflected electrons and holes in the presence of SOC. Furthermore, anomalous Andreev levels are observed for different HM length scales signifying its role in the tunability of the $\phi$ - phase Josephson junction. 

\end{abstract}

\pacs{74.45.+c, 85.75.-d, 74.90.+n, 75.76.+j}
\maketitle

\section{Introduction}
The exhibition of superconductivity and magnetism in the same domain leads to several engrossing phenomena \cite{saxena,aoki,pfleiderer,ginsburg,zutic,buzdin,acharjee1,hirai,andreev}. In general spin-singlet superconductivity has a detrimental effect with a strong magnetic field \cite{andreev,ginsburg,hirai}. Nevertheless, the discovery of long-range superconductivity in two-dimensional monolayer transition-metal dichalcogenides (TMD) like  MoS$_2$, NbSe$_2$, TaS$_2$, WSe$_2$ etc., open up a new dimension in superconducting spintronics \cite{barrera,idzuchi,jalouli,taguchi,scharf,tang,lu, saito, xi, dvir, costanzo, sohn, li, hamil, ai}. Monolayer TMD - superconductors lack in-plane mirror symmetry, resulting in strong antisymmetric spin-orbit coupling (SOC). So, they admit both spin-singlet and spin-triplet components of the superconducting wave function.
Moreover, due to the lack of inversion symmetry and strong SOC, the electrons near the $K$ and $-K$-valleys witness a strong Zeeman field more enormous than the Pauli limit, leading to long-range superconductivity \cite{lu, saito, xi, dvir, costanzo, sohn, li, hamil, ai,zhu2, xiao2,kormanyos,cappelluti}. It is to be noted that the Andreev reflection process is significantly suppressed in conventional HM$|$S heterostructure \cite{andreev,ginsburg,hirai}. However, half metal (HM)$|$Ising Superconductor (ISC) junctions can offer spin-triplet Andreev reflection due to the formation of equal spin Cooper pairs when the alignment of the HM magnetic moments is parallel to the plane of HM$|$ISC junction \cite{zhou,lv,cheng}. Furthermore, spin-triplet pairing will lead to nontrivial topological superconductivity supporting Majorana edge states \cite{zhou}. 

Thus, the magnetic heterostructure and Josephson junctions (JJ) have widely been explored theoretically and experimentally \cite{huy, flouquet,nandi,banerjee, acharjee2,acharjee3}. Efforts have been made to realize the coexistence of $0$ and $\pi$ states in the junction region and to fabricate a tunable $\phi$ - phase JJ, which can be used as a quantum phase battery \cite{strambini}. Moreover, different $\phi$ phases can be utilized to fabricate quantum qubit devices and cryogenic memories for next-generation quantum computing based on the currently available semiconductor technology \cite{ioffe,ioffe2,hilgenkamp,madden}. A $\phi$ - phase JJ can offer a constant phase bias $\phi_0$ in open circuit configuration resulting in the current phase relation (CPR), $J(\phi) = J_c\sin(\phi+\phi_0)$ \cite{buzdin2}. However, the closed superconducting loop will induce an anomalous Josephson current, $J(\phi) = J_c\sin(\phi_0)$. It is to be noted that $0$ or $\pi$ - phase JJ can be easily engineered if the time-reversal symmetry ($t \nrightarrow -t$) is broken \cite{strambini}. However, in the absence of both time reversal ($t \nrightarrow -t$) and inversion symmetries ($\mathbf{r} \nrightarrow -\mathbf{r}$), a finite phase shift $\phi_0$ in the CPR will be induced. Since, lateral hybrid arrangements break the inversion symmetry, they possess strong SOC. Thus lateral junctions with ISC or topological insulators are the promising candidate to engineer Josephson $\phi_0$ junction \cite{szombati,assouline,mayer}.

The JJs with spin-active interfaces or magnetic impurities can offer spin-sensitive Andreev reflection, supporting unconventional long-range spin-triplet correlations \cite{keizer1,khaire,linder1,eschrig,acharjee4}. The long-range spin-triplet superconducting correlation in such JJs is mainly due to the combination of spin mixing and spin-flip scattering processes \cite{eschrig2, galaktionov, eschrig,acharjee4, millis, hubler,kalenkov}. Experimentally, magnetic JJ with spin-active interfaces can be possibly realized by placing a thin magnetic layer at the interface. The barrier spin moments in such JJ are still uncontrollable as it is its inherent property. However, a well engineered spin-active interface can host and control the barrier spin moments through non-equilibrium spin injection process at the interface \cite{ouassou,zwierzycki,jeon,bobkova}. Supercurrent control and anomalous conductivities have been explored in superconductor-ferromagnetic geometries \cite{jeon}. Due to a spin-active barrier moment, the transmission profile of up-spin and down-spin electrons at the interface are significantly different and can be controlled through the spin-mixing angle. Moreover, the impact of spin-flip scattering on Josephson supercurrent and CPR has also been studied in a spin active S$|$HM$|$S  JJ \cite{galaktionov,kalenkov,millis}. Very recently, the $0 - \pi$ switch effect has been explored in ISC$|$FM JJs \cite{cheng}.
The transport properties of ISC$|$HM hybrids and JJs have also been explored theoretically \cite{dai}. However, from the best of our knowledge, the Ising Superconductor/half metallic heterostructure with spin active interfaces is yet to be explored. So, in this work, we consider an ISC$|$HM$|$ISC JJ with spin-active interfaces. The present work is focusses on understanding the superconducting spin-triplet correlations and the interplay between spin-dependent barrier moment, SOC of the ISC concerning (i) Andreev levels, (ii) Majorana modes and (iii) Josephson supercurrent.  

The paper's organization is as follows: Section II defines the Bogoliubov de-Gennes Hamiltonian and discusses a theoretical formulation of the proposed geometry. The Andreev Bound States (ABS) and the  supercurrent of the proposed system are studied in Section III. In Section IV, the results of our work are discussed. A brief note on realization of  the proposed experimental setup and concluding remarks of our work is presented in Section V. 

\section{Formulation}
\begin{figure}[hbt]
\centerline
\centerline{ 
\includegraphics[scale=0.6]{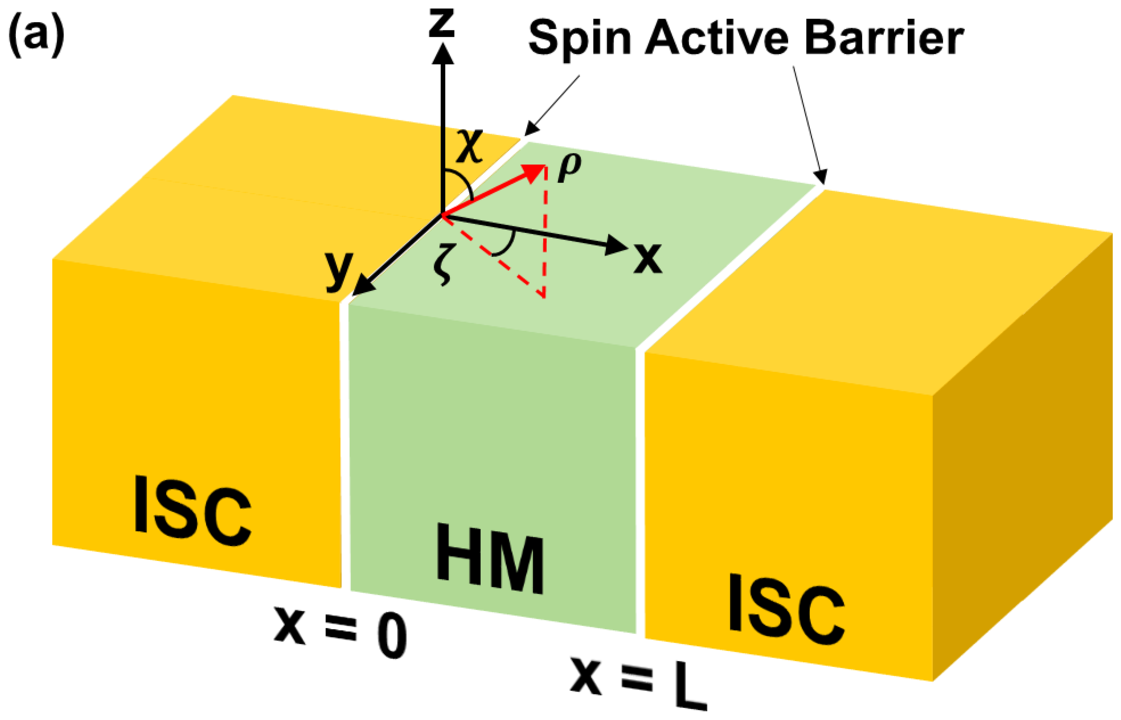}
\vspace{0.4cm}
\includegraphics[scale=0.4]{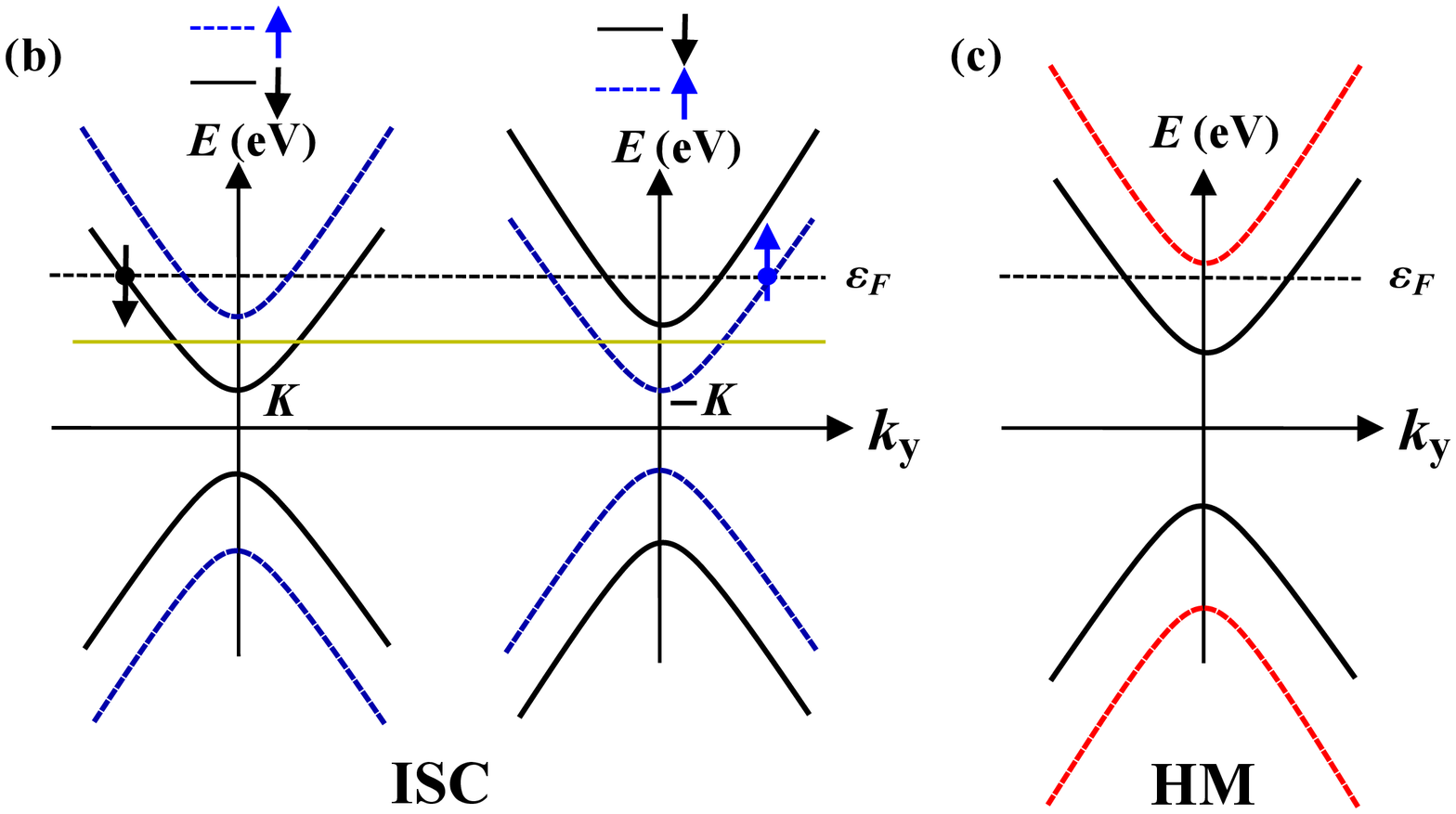}
\vspace{-0.5cm}
}
\caption{(a) Schematic illustration of an ISC$|$HM$|$ISC Josephson junction. We consider a half-metallic (HM) ferromagnet in strong proximity with the Ising superconductor (ISC). The interfaces are supposed to be located at $x = 0$ and $x = \text{L}$. We consider the HM's bulk magnetic moment $\mathbf{m}$ to be along the z-direction. The barrier magnetic moment $\rho$ is misaligned with bulk moments by polar angle $\chi$ and azimuthal angle $\zeta$. (b) Schematic energy band structure of an ISC near the $\pm\mathbf{K}$ valleys. The black dashed and green solid grey lines indicate the Fermi energy $\varepsilon_F$ of the double-band and single-band ISC, respectively. (c) Schematic energy band structure of a half-metallic ferromagnet where the Fermi energy of the HM is denoted by the black dashed line.}
\label{fig1}
\end{figure}

We consider a half-metallic (HM) ferromagnet sandwiched between two Ising Superconductors (ISC) as shown in Fig. \ref{fig1}(a). The ISC$|$HM$|$ISC Josephson junction is supposed to satisfy the condition $\xi << \text{L}$, where L is the length of the HM, and $\xi$ is the superconducting coherence length. 
The magnetization of the bulk HM is directed along the z direction, and the ISC$|$HM interface is spin active. So, the barrier magnetic moments constitute a spin-dependent potential $\mathcal{V}_{m}$ defined as $\mathcal{V}_{m} = \rho \mathcal{V}_0(\sin\chi\cos\zeta, \sin\chi\sin\zeta, \cos\chi)$, where, $\mathcal{V}_0$ is the intrinsic spin independent barrier potential. The parameter $\rho = |\mathcal{V}_{m}|/\mathcal{V}_0$ characterize the effective ratio of barrier magnetic and non-magnetic moments, and the angles $\chi$ and $\zeta$ represents the polar and the azimuthal angular components of the magnetic barrier moment made with the bulk moment of the HM.

Using Nambu basis, $\Psi = (\psi_\uparrow, \psi_\downarrow, \psi_\uparrow^\dagger, \psi_\downarrow^\dagger)$ the Bogoliubov de-Gennes (BdG) equation of the system can be written as $\check{\mathcal{H}}_\text{BdG}\Psi = \varepsilon\Psi$, where BdG Hamiltonian can be defined as \cite{acharjee4, cheng} 

\begin{equation}
\label{eq1}
\check{\mathcal{H}}_\text{BdG} = \left(
\begin{array}{cc}
 \hat{\mathcal{H}}_\pm(\mathbf{k}) +\hat{\mathcal{M}}  & \hat{\Delta}(\mathbf{k}) \\
 -\hat{\Delta }^{\ast}(\mathbf{k}) & -\hat{\mathcal{H}}^{\ast}_\pm(\mathbf{-k}) 
 -\hat{\mathcal{M}}^{\ast} \\
\end{array}\right)
\end{equation} 
\noindent where $\hat{\mathcal{H}}_\pm(\mathbf{k})$ is the single particle Hamiltonian which in the presence of the two valleys ($\mathbf{K}$ and $-\mathbf{K}$) of ISC can be written as \cite{tang}

\begin{equation}
\label{eq2}
\hat{\mathcal{H}}_\pm(\mathbf{k}) = \left(-\frac{\hbar^2k^2}{2m}-\mu\right)\hat{\text{I}} +\epsilon\beta\hat{\sigma}_z + \hat{\mathcal{V}}_{int}
\end{equation}
\noindent where $k$ is the wave vectors of the electron corresponding to the valley $\pm \mathbf{K}$, $\mu \equiv \mu_i[\Theta(-x) + \Theta(x-\text{L})]$ are the chemical potentials of the respective layers and $\epsilon = \pm$ is the valley index for $\pm \mathbf{K}$. The schematic energy band of the ISC is displayed in Fig. \ref{fig1}(b). The up and down spin subbands split into constituents due to the SOC of the ISC. It is to be noted that the spin-up band has higher energy than spin-down bands at $\mathbf{K}$ valley, while an opposite characteristic is observed for $-\mathbf{K}$ valley. The parameter $\beta$ characterize the strength of spin-orbit coupling of ISC while $\mathcal{\hat{V}}_{int}$ is the barrier potential at ISC$|$HM interfaces and can be defined as \cite{acharjee4,kalenkov}

\begin{equation}
\label{eq3}
\hat{\mathcal{V}}_{int} = (\mathcal{V}_0\hat{\text{I}}+\mathbf{\sigma}.\mathbf{\mathcal{V}}_m )\lbrace\delta(x)+\delta(x-\text{L})\rbrace
\end{equation}

The term $\hat{\mathcal{M}}$ in Eq. (\ref{eq1}) represents the exchange field of the system and can be defined as
\begin{equation}
\label{eq4}
\hat{\mathcal{M}} = \mathbf{m}.\mathbf{\sigma}\Theta(x)\Theta(\text{L}-x)
\end{equation}
\noindent where, the parameter, $\mathbf{m}$ represent the magnetization of the bulk HM and $\mathbf{\sigma}$ = ($\sigma_x,\sigma_y,\sigma_z$) are the Pauli's spin matrices. Since the spin polarization of the bulk HM is completely along the z-direction, we can write $\mathbf{m} \equiv (0,0,1)$. 

The superconducting pair potential for the left and right ISC appearing in Eq. (\ref{eq1}) can be defined as \cite{tang}
\begin{equation}
\label{eq5}
\hat{\Delta}(\mathbf{k}) = \Delta i \hat{\sigma}_y\lbrace e^{i\phi_\text{L}} \Theta(-x) + e^{i\phi_\text{R}} \Theta(x-\text{L})\rbrace,
\end{equation}
\noindent where $\Delta$ is the superconducting gap magnitude and $\phi_\text{L} (\phi_\text{R})$ correspond to the superconducting phase of the left (right) ISC. So, the phase difference $\phi$ of the left and right ISC can be written as $\phi = (\phi_\text{L} - \phi_\text{R})$. The wave function in the ISC regions can be obtained by considering $\hat{\mathcal{M}} = 0$ and solving the BdG equation.
Using plane wave assumption and diagonalizing the Hamiltonian from Eq.(\ref{eq1}), the wave function in the left ISC can be defined as \cite{cheng}

\begin{multline}
\label{eq6}
\Psi^{\text{L}}_{\text{ISC}\pm}(x<0) = 
a^\text{L}_\pm\left(u e^{i\phi_{\text{L}}/2}\hat{\varphi}_1 + v e^{-i\phi_{\text{L}}/2}\hat{\varphi}_4\right)e^{-ik_\pm x}\\
+b^\text{L}_\pm\left(u e^{i\phi_{\text{L}}/2}\hat{\varphi}_2 - v e^{-i\phi_{\text{L}}/2}\hat{\varphi}_3\right)e^{-ik_\mp x}\\+
c^\text{L}_\pm\left(v e^{i\phi_{\text{L}}/2}\hat{\varphi}_1 
+ u e^{-i\phi_{\text{L}}/2}\hat{\varphi}_4\right)e^{ik_\pm x}\\
+d^\text{L}_\pm\left(-v e^{i\phi_{\text{L}}/2}\hat{\varphi}_2 
+u e^{-i\phi_{\text{L}}/2}\hat{\varphi}_3\right)e^{ik_\mp x}
\end{multline}
Similarly, the wave function of the right ISC can also be defined as
\begin{multline}
\label{eq7}
\Psi^{\text{R}}_{\text{ISC}\pm}(x>L) = 
a^\text{R}_\pm\left(u e^{i\phi_{\text{R}}/2}\hat{\varphi}_1 + v e^{-i\phi_{\text{R}}/2}\hat{\varphi}_4\right)e^{ik_\pm x}\\
+b^\text{R}_\pm\left(u e^{i\phi_{\text{R}}/2}\hat{\varphi}_2 - v e^{-i\phi_{\text{R}}/2}\hat{\varphi}_3\right)e^{ik_\mp x}\\+
c^\text{R}_\pm\left(v e^{i\phi_{\text{R}}/2}\hat{\varphi}_1 
+ u e^{-i\phi_{\text{R}}/2}\hat{\varphi}_4\right)e^{-ik_\pm x}\\
+d^\text{R}_\pm\left(-v e^{i\phi_{\text{R}}/2}\hat{\varphi}_2 
+u e^{-i\phi_{\text{R}}/2}\hat{\varphi}_3\right)e^{-ik_\mp x}
\end{multline}
\noindent where we define, $\hat{\varphi}_1 = (1,0,0,0)^\text{T}$, $\hat{\varphi}_2 = (0,1,0,0)^\text{T}$, 
$\hat{\varphi}_3 = (0,0,1,0)^\text{T}$ and
 $\hat{\varphi}_4 = (0,0,0,1)^\text{T}$. Here,  $a^\text{L}_\pm$ ($b^\text{L}_\pm$) are the reflection 
 coefficients for up (down) spin electrons  while  $c^\text{L}_\pm$ ($d^\text{L}_\pm$) are the reflection coefficients for the up (down) spin holes in left ISC region. $a^\text{R}_\pm$ ($b^\text{R}_\pm$) are the transmission coefficients for up(down) spin electrons while   $c^\text{R}_\pm$ ($d^\text{R}_\pm$) are the transmission coefficients for the up(down) spin holes in right ISC region. $k_{+(-)}$ gives the momenta of the electron (hole) in ISC. The momenta of the electron and the holes under the Andreev approximation can be defined as 
 \begin{equation}
\label{eq8}
k_{\pm} = \sqrt{\left(2\mu_\text{S} \mp \beta\right)}
\end{equation}
where $\mu_\text{S}$ and $\beta$, respectively, are the chemical potential and the strength of SOC in the ISC region. We also have considered $m = \hbar = 1$ for our analysis. The quasiparticle amplitudes $u$ and $v$ appearing in Eq.(\ref{eq8}) are defined as
\begin{eqnarray}
\label{eq9}
\label{eq10}
 u =  \frac{1}{\sqrt{2}}\sqrt{1+\sqrt{1-\frac{\Delta^2}{E^2}}}\\
 v =  \frac{1}{\sqrt{2}}\sqrt{1-\sqrt{1-\frac{\Delta^2}{E^2}}}
\end{eqnarray}
 In a similar way, the wave function in the half-metallic region is \cite{linder1,acharjee4}
\begin{multline}
\label{eq11}
 \Psi_{\text{HM}\pm}(0 < x < L) = e_\pm\hat{\varphi}_1e^{i\kappa^+_{+}x} + f_\pm\hat{\varphi}_2e^{i\kappa^+_{-}x} 
\\ + g_\pm\hat{\varphi}_3e^{-i\kappa^-_{+}x} + h_\pm\hat{\varphi}_4e^{-i\kappa^-_{-}x}
\end{multline}
\noindent where $e_\pm$, $f_\pm$, $g_\pm$, $h_\pm$ are the scattering coefficients for electron and hole in the half-metallic region with $\pm$ corresponding to the wave function $ \Psi_{\text{HM}\pm}$ respectively. The quasi-particle momenta in this region are given by 
\begin{equation}
\label{eq12}
\kappa^\pm_{\sigma} =\pm\mu_\text{HM}\mp\sqrt{(m_z+\mathcal{V}_{mz})^2+\mathcal{V}_{mx}^2+\mathcal{V}_{my}^2}
\end{equation}
\noindent where, $\mu_\text{HM}$ is the chemical potential of the HM and ($\mathcal{V}_{mx}$, $\mathcal{V}_{my}$, $\mathcal{V}_{mz}$) are the components of barrier magnetic moments. The schematic energy band of the HM is displayed in Fig. \ref{fig1}(c). The spin-up and spin-down subbands split into constituents due to the strong Zeeman field of the HM. The wave functions must satisfy the conditions:
\begin{align}
\label{eq13}
\Psi^{\text{L}}_{\text{ISC}\pm}(0^-) &= \Psi_{\text{HM}\pm}(0^+),\\
\label{eq14}
\Psi_{\text{HM}\pm}(\text{L}^-) &= \Psi^{\text{R}}_{\text{ISC}\pm}(\text{L}^+),\\
\label{eq15}
\Psi^{'}_{\text{HM}\pm}(0^+)-\Psi^{\text{L}'}_{\text{ISC}\pm}(0^-)
&= \hat{\eta}\Psi^{\text{L}}_{\text{ISC}\pm}(0^-),\\
\label{eq16}
\Psi^{\text{R}'}_{\text{ISC}\pm}(\text{L}^+)-\Psi^{'}_{\text{HM}\pm}(\text{L}^-)
&= \hat{\eta}\Psi_{\text{HM}\pm}(\text{L}^-)
\end{align}
\noindent where, $\hat{\text{I}}$ is a 4$\times$4 identity matrix. The matrix $\hat{\eta}$ gives the sum of spin-independent and spin-dependent barrier potential and has the following form \cite{linder1,acharjee4}
 \begin{equation}
\label{eq17}
\hat{\eta} = \left(
\begin{array}{cccc}
Z + \Omega_1 & \Omega_2 & 0 & 0 \\
\Omega_2^* &  Z -\Omega_1 & 0 & 0 \\
0 &  0 & Z + \Omega_1& \Omega_2^* \\
 0 &  0 & \Omega_2 & Z -\Omega_1 \\
\end{array}
\right)
\end{equation}
where, $Z=\frac{2\mathcal{V}_0}{k_F\cos\theta}$ represents the spin independent barrier strength while $\Omega_1 = 2\rho \mathcal{V}_0\cos\chi$ and $\Omega_2 = 2\rho \mathcal{V}_0\sin\chi e^{-i\zeta}$ are the terms corresponds to spin dependent barrier potential.

\begin{figure*}[hbt]
\centerline
\centerline{
\includegraphics[scale = 0.52]{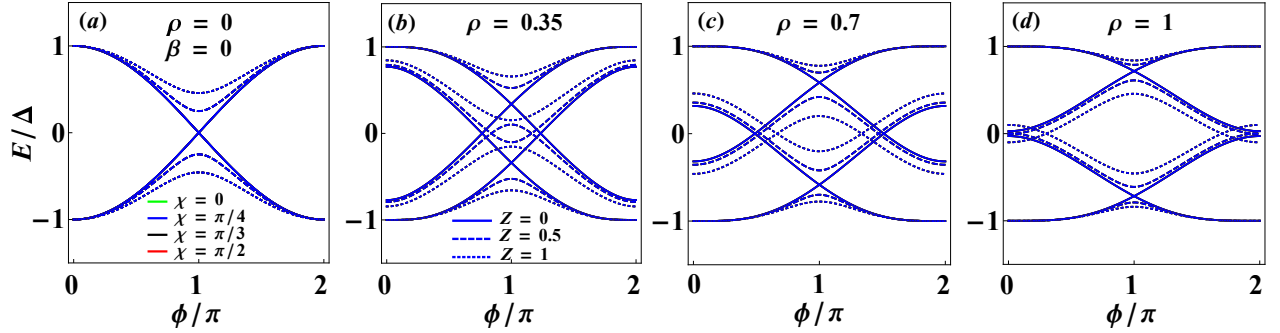}}
\caption{Andreev levels ($E/\Delta$) as a function of phase difference ($\phi$) for various values of $\rho$ in absence of SOC ($\beta = 0$) considering $\text{L}/\text{L}_0 = 0.01$. The ABS spectra are plotted for $Z = 0$ (solid blue line), $Z = 0.5$ (dashed blue line) and $Z = 1$ (dotted blue line).
}
\label{fig2}
\end{figure*} 

\section{Bound State energy and Supercurrent}
The Andreev levels of the ISC$|$HM$|$ISC Josephson junction can be obtained by using the boundary conditions  Eqs. (\ref{eq13})-(\ref{eq16}). This indeed gives rise to a system of equations given by $\hat{\mathcal{A}}\hat{x} = 0$, where $\hat{\mathcal{A}}$ is a 16$\times$16 matrix carrying all the information regarding the bound state energies and $\hat{x} = (a^\text{L}_\pm,b^\text{L}_\pm,c^\text{L}_\pm,d^\text{L}_\pm,e_\pm,f_\pm,g_\pm,h_\pm,e_\pm',
f_\pm',g_\pm',h_\pm',a^\text{R}_\pm,b^\text{R}_\pm,c^\text{R}_\pm,$ 
$d^\text{R}_\pm)^\text{T}$. The bound state energies and the Andreev levels are obtained by using the condition $det[\hat{\mathcal{A}}(E_\pm)] = 0$. where, $det[...]$ represent the determinent of the matrix.
An extensive form of $\hat{\mathcal{A}}(E_\pm)$ is presented in Appendix (\ref{A1}). It is to be noted that the Andreev levels satisfy the equality $E_- = -E_+$ \cite{cheng}. So, in our analysis, only the results of Andreev levels $E_+$ are displayed.

An expression for the normalized Josephson supercurrent can be obtained by using the standard expression \cite{linder1,acharjee4}
\begin{equation}
\label{eq18}
\frac{J}{J_0} = \int_{-\frac{\pi}{2}}^{\frac{\pi}{2}} d\theta\sum_\sigma \tanh\left(\frac{ E_{\sigma}}{2 k_B T}\right) \frac{dE_\sigma}{d\phi}
\end{equation}  
where, $J_0 = 2e\Delta/\hbar$ and $E_\sigma$ corresponds to the energies of the Andreev levels with $\sigma = \pm 1$. Here, $\theta$ is the angle of incidence of the incoming electron. We set, $T/T_c = 0.001$  with $T_c$ represents the critical temperature of the ISC and $\mu_\text{S} = 1$. Moreover, to restrict our analysis to the interface, we consider the azimuthal angle $\zeta = \frac{\pi}{2}$ for all our analyses. The supercurrent generally has $k_F$ dependence for large values of $Z$. So, we consider both low ($Z = 0$) and high ($Z = 1$) values of the barrier width. In our analysis, the spin mixing processes are achieved by considering $\rho \neq 0$ and $\chi = 0$ while spin mixing and spin flipping processes are obtained for $\rho \neq 0$ and $\chi \neq 0$.
 
\begin{figure*}[hbt]
\centerline
\centerline{
\includegraphics[scale = 0.52]{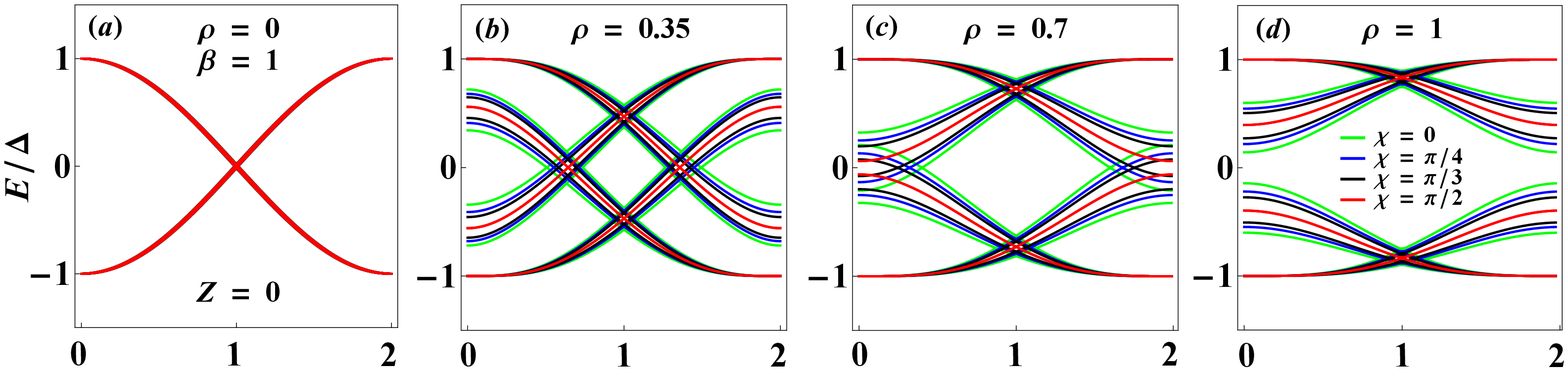}
\includegraphics[scale = 0.52]{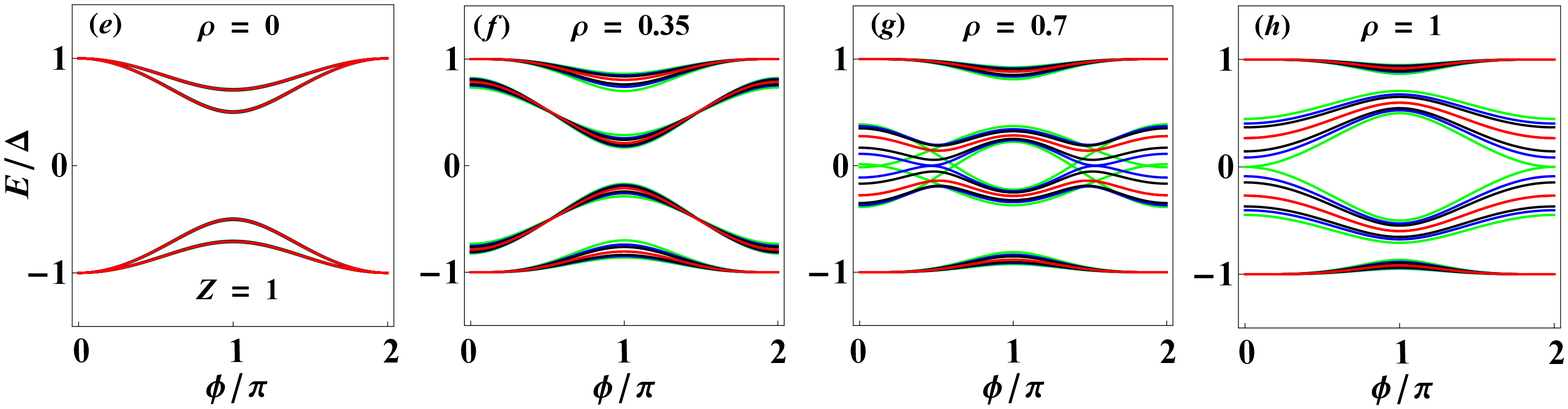}}
\caption{Andreev levels ($E/\Delta$) as a function of phase difference ($\phi$) for various values of $\rho$ considering $\text{L}/\text{L}_0 = 0.01$. The plots in the top and middle panels represent the ABS spectra for different choices of $\chi$ with $Z = 0$ and $Z = 1$, respectively, considering $\beta = 1$. The Andreev levels are plotted for the mismatch angle $\chi = 0$ (green solid line), $\chi = \frac{\pi}{4}$ (solid blue line), $\chi = \frac{\pi}{3}$ (solid black line) and $\chi = \frac{\pi}{2}$ (solid red line).}
\label{fig3}
\end{figure*}
\section{Andreev Bound States}
\subsection{Andreev levels for a spin active barrier in absence of Spin-orbit coupling}
 Before we present the result of 
a spin-active ISC$|$HM$|$ISC JJ, we have presented the Andreev energy levels in the absence of SOC which indeed represents a spin-active triplet S$|$HM$|$S JJ. We have considered different $\rho$ and $\chi$ values to understand the role of spin mixing and spin flipping on ABS spectra. In the absence of SOC, the Andreev levels are expressed by
\begin{multline}
\label{eq19}
\frac{E_\pm}{\Delta}  = \pm\sqrt{\frac{\mathcal{P}_1+\mathcal{P}_2 \cos \phi \pm 4 \rho \sqrt{\mathcal{P}_3-\mathcal{P}_4\cos \phi-\cos 2 \phi}}{\left(Z^2+2\right)^2-8 \rho ^2 \left(Z^2-2\right)+16 \rho^4}}
\end{multline}
where, we define $\mathcal{P}_1$, $\mathcal{P}_2$, $\mathcal{P}_3$ and $\mathcal{P}_4$ as
\begin{align}
\mathcal{P}_1 &= 16 \rho ^4+4 \rho ^2+Z^4+\left(3-8 \rho ^2\right) Z^2+2,\nonumber\\ 
\mathcal{P}_2 &=-4 \rho ^2+Z^2+2, \nonumber\\
\mathcal{P}_3 &= 8 \rho ^2+2 Z^2+1,\nonumber\\ 
\mathcal{P}_4 &= 2 \left(Z^2-4 \rho ^2\right),\nonumber
\end{align}
The Andreev energy levels from Eq. (\ref{eq19}), are displayed in Fig. (\ref{fig2}) for different barrier width ($Z$), barrier magnetic moment ($\rho$) and mismatch angle ($\chi$). The ABS spectra are found to be degenerate for $\rho = 0$, which follows the results of a spin inactive conventional S$|$F$|$S JJ. A splitting of the two degenerate energy bands into four respective constituents is observed for $\rho = 0.35$, which arises solely due to the spin-mixing process in a spin-active barrier. However, the identical ABS spectra for different $\chi$ values indicate no impact of spin-flipping processes on Andreev levels. A signature of non-trivial Josephson phases $\phi_1$ and $\phi_2$ are observed as minimum energy for $\phi$ state in transparent ($Z = 0$) and intermediate ($Z = 0.5$) barrier width. However, the system favours the $\pi$ phase for an opaque barrier. The ($\phi_1$, $\phi_2$) states are mostly located near $\pi$ - phase. However, as $\rho$  increase to $0.7$, the  ($\phi_1$,  $\phi_2$) - phases depart from $\pi$, indicating the increase in the splitting of the Andreev levels. Moreover, in this configuration, the system resides in the $\phi$ phase for considered $Z$ values, as seen from Figs. \ref{fig2}(b) and \ref{fig2}(c). As $\rho$ increases to $1$, the two sets of Andreev levels depart further away from each other and tend to favour $0$ and $2\pi$ phases. Thus for a fully magnetic and transparent barrier, one can observe $0$ and $2\pi$ junctions. However, for a fully opaque barrier, the signature of $\phi$ junction is perceived from Fig. \ref{fig2}(d). Thus the results indicate the possibility of tunability via spin active barrier moment even in the absence of SOC. It is to be noted that the impact of spin flipping is absent for $\beta = 0$. Nevertheless, a possible Majorana mode can exist in a near-transparent barrier configuration due to the barrier spin moments \cite{tang}.

\subsection{Andreev levels for a spin active barrier in presence of Spin-orbit coupling}
In presence of SOC, the wave vectors $k_+$, and $k_-$ are not identical. This mismatch can cause additional splitting of the Andreev energy bands. It is to be noted that an analytic expression for Andreev levels for $\beta = 1$ is too challenging to obtain as both $k_+$ and $k_-$ has non-vanishing distinct values in this condition. So, we have used numerical simulation to present our result for $\beta = 1$. However, a possible analytic expression for a system with $\beta = 2$ can still be obtained as $k_-$  vanishes in this condition. The Andreev energy levels for $\beta = 2$ can be expressed as

\begin{widetext}
\begin{eqnarray}
\label{eq20}
\frac{E_\pm}{\Delta} = \pm \sqrt{\frac{\mathcal{R}_1+4\mathcal{R}_2\rho ^2\cos 2\chi+16\rho^4 \cos 4\chi\pm 8\sqrt{2\lbrace 2\mathcal{R}_3 \cos\phi+\mathcal{R}_4(\mathcal{R}_5-\lambda_3 \mathcal{R}_6 \cos \phi-2 \mathcal{R}_6^2 \cos ^2\phi)\rbrace}}{\lambda_1+16\lambda_2\cos 2\chi+32 \rho^4 \cos4\chi}}
\end{eqnarray}
\end{widetext}
where, we defined
\begin{align}
\lambda_1 &= \mathcal{S}_1-16 \rho ^2 \left(Z^4+3 Z^2+4\right) Z^2+\left(Z^2+4\right)^2 Z^4, \nonumber\\
\lambda_2 &= 8 \left(2 \rho ^4+\rho ^2\right)+Z^4-4 \left(2 \rho ^2+1\right) Z^2, \nonumber\\
\lambda_3 &= \left(Z^2-4 \rho ^2\right)^2, \nonumber\\
\mathcal{R}_1&= \mathcal{S}_2-4 \rho ^2 \left(4 Z^4+11 Z^2+8\right) Z^2+\left(Z^4+6 Z^2+8\right) Z^4, \nonumber\\
\mathcal{R}_2&=16 \left(\rho ^4+\rho ^2\right)+Z^4-8 \left(\rho ^2+1\right) Z^2, \nonumber\\
\mathcal{R}_3&=8 \rho ^4 \cos (4 \chi )-2 \rho ^2 \mathcal{R}_7 \cos (2 \chi )+\mathcal{S}_3, \nonumber\\
\mathcal{R}_4&= \rho^2 \lambda_3\cos^2\chi, \nonumber\\
\mathcal{R}_5&=4 \rho ^4 \left(8 \rho ^2+3\right)+Z^6+\left(2-6 \rho ^2\right) Z^4-8 \rho ^2 Z^2, \nonumber\\
\mathcal{R}_6&=-2 \rho ^2 \cos (2 \chi )-2 \rho ^2+Z^2, \nonumber\\
\mathcal{R}_7&=16 \rho ^4-8 \rho ^2 \left(Z^2+2\right)+Z^2 \left(Z^2+8\right), \nonumber\\
\mathcal{S}_1 &=256 \rho ^8+96 \rho ^4 \left(Z^4+1\right)-256 \rho ^6 \left(Z^2-1\right), \nonumber\\
\mathcal{S}_2 &=256 \rho ^8+64 \rho ^6 \left(1-4 Z^2\right)+16 \rho ^4 \left(6 Z^4+4 Z^2+3\right), \nonumber\\
\mathcal{S}_3 &=-32 \rho ^6+8 \rho ^4 \left(4 Z^2+3\right)-2 \rho ^2 Z^2 \left(5 Z^2+8\right)
+\mathcal{S}_4, \nonumber\\
\mathcal{S}_4 &= Z^4 \left(Z^2+4\right), \nonumber
\end{align} 

\begin{figure*}[hbt]
\centerline
\centerline{
\includegraphics[scale = 0.52]{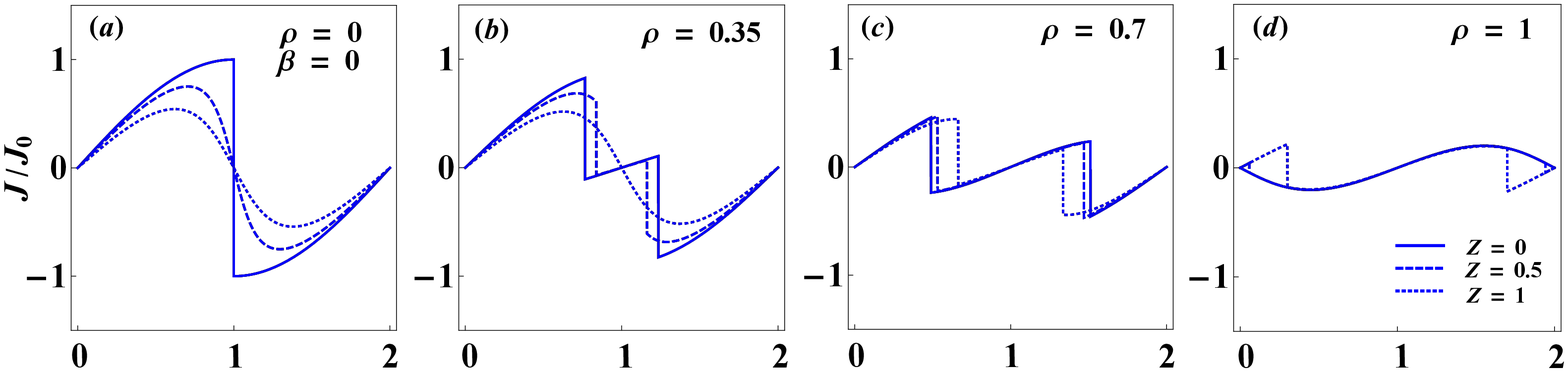}
\includegraphics[scale = 0.52]{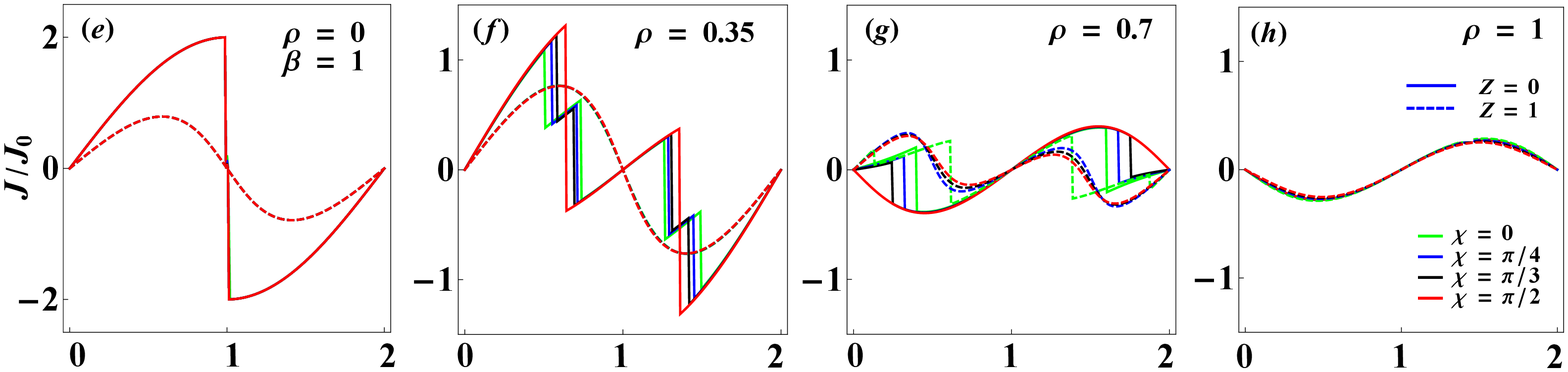}
\includegraphics[scale = 0.52]{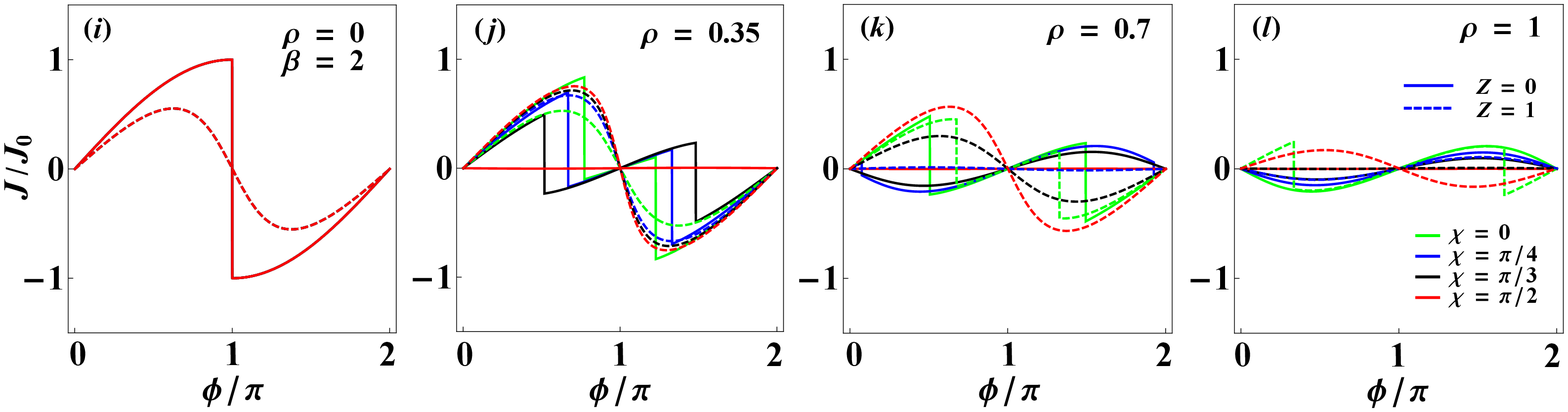}}
\caption{Josephson supercurrent ($J/J_0$) as a function of phase difference ($\phi$) for various values of $\rho$ and $Z$ with $\beta = 0$ (top panel) and $\beta = 2$ (middle panel) considering $\text{L}/\text{L}_0 = 0.01$. The plots in the middle panel are for the polar mismatch angle $\chi = 0$ (solid green line), $\chi = \frac{\pi}{4}$ (solid blue line), $\chi = \frac{\pi}{3}$ (solid black line) and $\chi = \frac{\pi}{2}$ (solid red line) respectively. The plots in the bottom panel are for $Z = 0$ and $\chi = \frac{\pi}{4}$ considering $\beta = 0$ (solid blue line), $\beta = 0.5$ (solid green line) and $\beta = 1$ (solid black line).
}
\label{fig4}
\end{figure*} 

Figs. \ref{fig3}(a) - \ref{fig3}(h) display the impact of barrier magnetic moment ($\rho$) and mismatch angle ($\chi$) on ABS spectra for transparent $(Z = 0)$ and opaque $(Z = 1)$ barrier configuration considering $\beta = 1$. The characteristics of the Andreev levels are similar with $\beta = 0$ seen from Figs. \ref{fig2}(a) for transparent barrier configuration.  The ABS bands are found to be degenerate in this case for $\rho = 0$ as seen from Fig. \ref{fig3}(a). However, a splitting in ABS spectra is observed in Fig. \ref{fig3}(e) for opaque barrier configuration, which arises due to the inherent SOC of the ISC.  With the rise in $\rho$ to $0.35$, the degenerate Andreev bands split into eight constituents due to the combined effect of SOC and spin active barrier moment. It is to be noted that the $\phi_1$ and $\phi_2$ phases individually split into two different $\phi$-states corresponding to the minima of the ABS spectra as indicated by Fig. \ref{fig3}(b). It is a consequence of different spin-flipping probabilities of the spin-up and spin-down electrons, which are unlocked due to SOC.  
So, it is possible to tune different $\phi$ states by considering suitable mismatch angles between the barrier and bulk moments. However, Andreev levels remain degenerate for $\chi = \pi/2$, indicating equal spin flipping probability of up and down spin electrons. The Andreev bands split further with the increase in $\rho$ to $0.7$. It is to be noted that there exists a possible $2\pi$ phase shift for $\chi = \pi/2$, indicating a possible $2\pi$ junction. But a $\phi$ - phase JJ is observed for $0\leq\chi<\pi/2$. 
Moreover, for a fully magnetic barrier, i.e., $\rho = 1$, the system prefers $2\pi$  junction for all mismatch angles in both transparent and opaque barrier configurations, as seen from Figs. \ref{fig3}(d) and \ref{fig3}(h) respectively.  Though splitting of the ABS bands is also observed with the increase in $\rho$ for $Z = 1$, but a notable difference in ABS spectra is noticed for $\rho = 0.35$. In this case, the system prefers the $\pi$ phase for all choices of $\chi$, as seen from Fig. \ref{fig3}(f). The splitting is further increased for $\rho = 0.7$; in this condition, the system resides in $\phi$ state as seen from Fig. \ref{fig3}(g). Furthermore, the existence of $\pi$-modes and the combined effect of barrier magnetic moment, mismatch angle and the SOC of the ISC can support Majorana mode \cite{acharjee4}. So, it can be concluded that the tunability of $0-\pi$ junction can be achieved via SOC of ISC, magnetic barrier moment and spin mismatch angle.

\begin{figure}[hbt]
\centerline
\centerline{
\includegraphics[scale = 0.47]{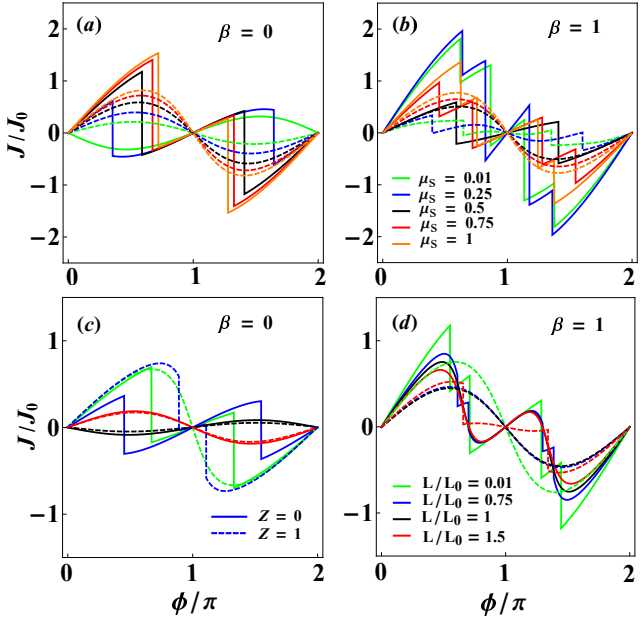}}
\caption{Josephson supercurrent ($J/J_0$) as a function of phase difference ($\phi$) for different values of $\rho$ and $Z$ with $\beta = 0$ and $\chi = \frac{\pi}{4}$ considering $\text{L}/\text{L}_0 = 0.01$ (solid green line), $\text{L}/\text{L}_0 = 0.75$ (solid blue lines),  $\text{L}/\text{L}_0 = 1$ (solid black lines) and $\text{L}/\text{L}_0 = 1.5$ (solid red lines).}
\label{fig5}
\end{figure}

\subsection{Current Phase Relation}
The current phase relation (CPR) of ISC$|$HM$|$ISC JJ are shown in Fig. \ref{fig4} corresponding to the ABS spectra in Figs. \ref{fig2} and \ref{fig3}. The Josephson supercurrents ($J/J_0$) are found to be similar to a conventional magnetic S/F/S junction in the absence of SOC \cite{linder1, acharjee4}, as seen from Fig. \ref{fig4}(a). A signature of anomalous supercurrent is noted from Figs. \ref{fig4}(b)-\ref{fig4}(d) for $\beta = 0$ and $\rho>0$. The system behaves like a $\phi$-junction for $Z = 0$ and $Z = 0.5$ with $\rho = 0.35$. However, the characteristics of a $\pi$-junction are still present for an opaque barrier ($Z = 1$). The further increase in $\rho$ to $0.7$ restrict the system to reside only in $\phi$-phase as indicated by Fig. \ref{fig4}(c). For a system with $\rho = 1$ and $Z = 0$, a phase shift of $\pi$ is noted. Though the system tends to behave like a $2\pi$ junction, the signature of $\phi$-state is still present for a system with barrier width $Z = 0.5$ and $Z = 1$ as seen from Fig. \ref{fig4}(d). So, it can be concluded that a $\phi$-junction can be realized by choosing a suitable barrier width and barrier magnetic moment,  even in the absence of SOC. Although the system is spin-active, the impact of $\chi$ is absent in the absence of SOC.

The plots in the middle and the bottom panel of Fig. \ref{fig4} display the CPR for $\beta = 1$ and $\beta = 2$, respectively. For a normal barrier with $\beta = 1$ and $\beta = 2$, the supercurrent is found to be similar to $\beta = 0$, which is consistent with the ABS spectra in Fig. \ref{fig3}(a) and \ref{fig3}(e). However, the magnitude of $J/J_0$ decreases for $\beta = 2$ due to the presence of $k_-$ component of the momenta. We observe the characteristics of a $\phi$-junction for a system with $Z = 0$, $\rho = 0.35$,  and $\beta = \mu_\text{S} = 1$. As $\phi_1$ and $\phi_2$ phases of the ABS spectra differ for different mismatch angles [Fig. \ref{fig3}(b)], consequently different $\phi$ junctions are noted in the CPR.  However, a $\pi$ - junction signature is observed for a system with $Z = 1$. Moreover, the impact of mismatch angle is absent in this case, as seen from Fig. \ref{fig4}(f). For $\rho = 0.7$ with $Z = 0$, a transition from $0 - \pi$ junction is noticed for mismatch angle $\chi = \frac{\pi}{2}$. While for other mismatch angles, the system still prefers the $\phi$ - phases as indicated by Fig. \ref{fig4}(g). A possible $\frac{\pi}{2}$ junction is noticed for an opaque barrier ($Z = 1$). The effect of the mismatch angle is suppressed, and a phase-shift of $\pi$ is noticed in $\rho = 1$ for both $Z = 0$ and $Z = 1$ conditions indicated from Fig. \ref{fig4}(h). The effect of spin mismatch angle is enhanced for $\beta = 2$ with $\rho \neq 0$. However, in the presence of spin magnetic moment with $\chi=\frac{\pi}{2}$, the effect of spin-flipping is absent, resulting in a vanishing supercurrent. A signature of anomalous supercurrent is noticed in Figs. \ref{fig4}(j) - \ref{fig4}(l) for $\rho = 0.7$ and $\rho = 1$, which is due to the presence of spin mixing and spin flipping processes for both transparent and opaque barrier conditions. Moreover, the presence of SOC and spin flipping enhances the tunability of $\phi$-phase even in the opaque barrier. This indicates that ISC$|$HM$|$ ISC-based JJ can be utilized to fabricate a tunable $\phi$ - junction more efficiently than the conventional JJs.

Fig. \ref{fig5}, display a comparative result of CPR between conventional spin-active S$|$HM$|$S JJ and a spin-active ISC$|$HM$|$ISC JJ considering $\rho = 0.35$ and $\chi = 0.25\pi$. The impact of chemical potential ($\mu_\text{S}$) is shown in Figs. \ref{fig5}(a) and \ref{fig5}(b) considering $\text{L}/\text{L}_0 = 0.01$ while that of the length of the HM ($\text{L}/\text{L}_0$) is shown in Figs. \ref{fig5}(c) and \ref{fig5}(d) considering $\mu_\text{S} = 1$. For a system with $\mu_\text{S} << 0.25$, a possible $2\pi$ - junction is observed while the system favours $\phi$-phase for $0.25 \leq\mu_\text{S} \leq 1$ for $\beta = 0$ and $Z = 0$ which is in accordance with the results of Ref[\cite{linder1}. However, in case of $\beta = 1$ with $Z = 0$, the system tends to form $\phi$- junction for $0.01 \leq\mu_\text{S} \leq 1$. In case of opaque barrier ($Z = 1$), a conventional spin-active JJ mostly reside in $\pi$-phase while a spin-active ISC$|$HM$|$ISC JJ favours $\pi$-phase only for $0.5 \leq\mu_\text{S} \leq 1$. In presence of SOC an anomalous $\phi$-junction is observed for $0.01 \leq\mu_\text{S} < 0.5$. Furthermore, apart from conventional $\phi$- characteristics, additional spikes in CPR are observed for a spin-active ISC$|$HM$|$ISC JJ in transparent barrier condition. It is due to the additional splitting of the bands solely by the inherent SOC leading to the mismatch of $k_+$ and $k_-$ vectors of ISC. It is noteworthy to mention that the system transits from $2\pi$ to $\pi$ phase for both transparent and opaque conventional spin-active S$|$HM$|$S JJ as the $\text{L}/\text{L}_0$ is increased from $1$ to $1.5$. However, a possible $\phi$-junction is observed for all choices of $\text{L}/\text{L}_0$ in a spin-active ISC$|$HM$|$ISC JJ.

\subsection{Andreev reflection and Tunnelling Conductance}
The tunneling conductance in presence of SOC, spin mixing and spin flipping can be obtained by using the relation \cite{hashimoto,acharjee2,acharjee3,dai}
\begin{equation}
\label{eq21}
\frac{\sigma}{\sigma_0} = \sum_{i = 1,2} \left(1 - |b^i_\pm|^2 + |d^i_\pm|^2\right)
\end{equation}
where, $b^i_\pm$ and $d^i_\pm$ are the normal and Andreev reflection probabilities at the respective layers.

\begin{figure}[hbt]
\centerline
\centerline{
\includegraphics[scale = 0.43]{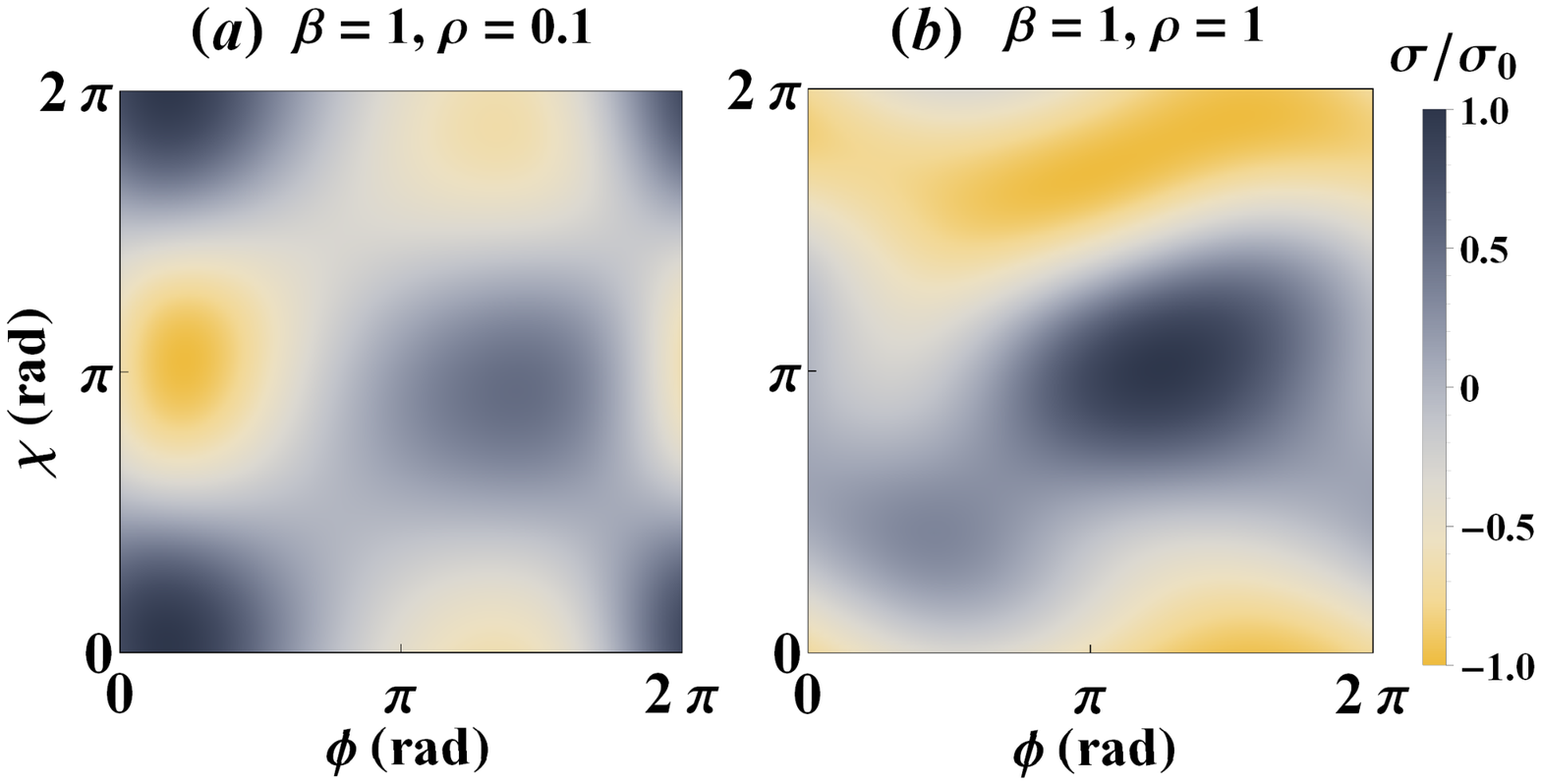}
\includegraphics[scale = 0.4]{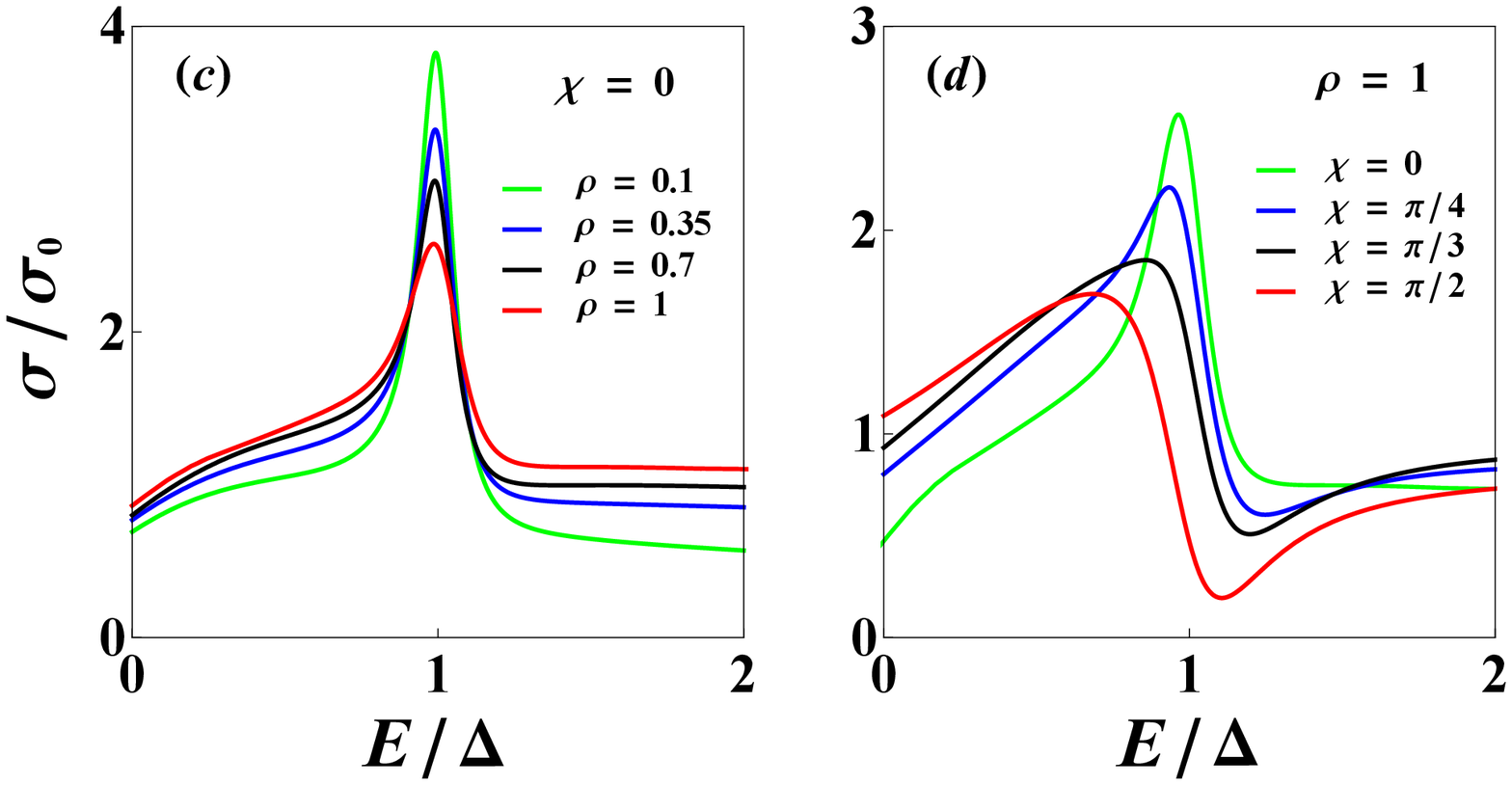}}
\caption{(Top panel) Density plot of the conductance variation with $\chi$ and $\phi$ for different barrier magnetic moments. (Bottom panel) Tunnelling conductance spectra for (c) $\rho = 0.1$ (solid green line), $\rho = 0.35$ (solid blue line), $\rho = 0.7$ (solid black line) and $\rho = 1$ (solid red line) with $\chi = 0$ while plot (d) corresponds to conductance for $\chi = 0$ (solid green line), $\chi = \frac{\pi}{4}$ (solid blue line), $\chi = \frac{\pi}{3}$ (solid black line) and $\chi = \frac{\pi}{2}$ (solid red line) with $\rho = 1$.}
\label{fig6}
\end{figure}
The tunnelling conductance of ISC$|$HM$|$ISC JJ is shown in Fig.  \ref{fig6}. The orientation ($\chi$) and phase ($\phi$) dependence of the conductance profile are displayed for $\rho = 0.1$ and $1$ in Figs. \ref{fig6}(a) and \ref{fig6}(b) respectively considering $\text{L}/\text{L}_0 = 0.01$, $\beta = 1$ and $E/\Delta = 0.5$. An asymmetric characteristic of the conductance spectra is observed for strong magnetic barriers in contrast to a low magnetic barrier due to the combined effect of SOC, spin mixing and spin flipping processes. It is to be noted that for a strong magnetic barrier, there exists a sudden dip in conductance profile in the region $\frac{3\pi}{2} <\chi \leq 2\pi$. The variation of conductance with energy ($E$) in terms of gap parameter ($\Delta$) for spin mixing processes is shown in Fig. \ref{eq7}(c). At the same time, we consider a combined effect of spin mixing and spin flipping on conductance in Fig. \ref{fig6} (d). We observe that the sharpness of the peak decreases with an increasing value of $\rho$. Although the sub-gap conductance is zero for conventional S$|$HM$|$S JJ's, a finite sub-gap conductance is observed in our proposed system. It is due to unequal probabilities of Andreev reflected electron and hole due to SOC. In the presence of spin-flipping processes with $\rho = 1$, the zero bias conductance peaks are found to be different for different mismatch angles. Moreover, the conductance peaks broaden and shift towards the low energy values with an increased mismatch angle, as shown in Fig. \ref{fig6}(d).

\begin{figure}[hbt]
\centerline
\centerline{
\includegraphics[scale = 0.48]{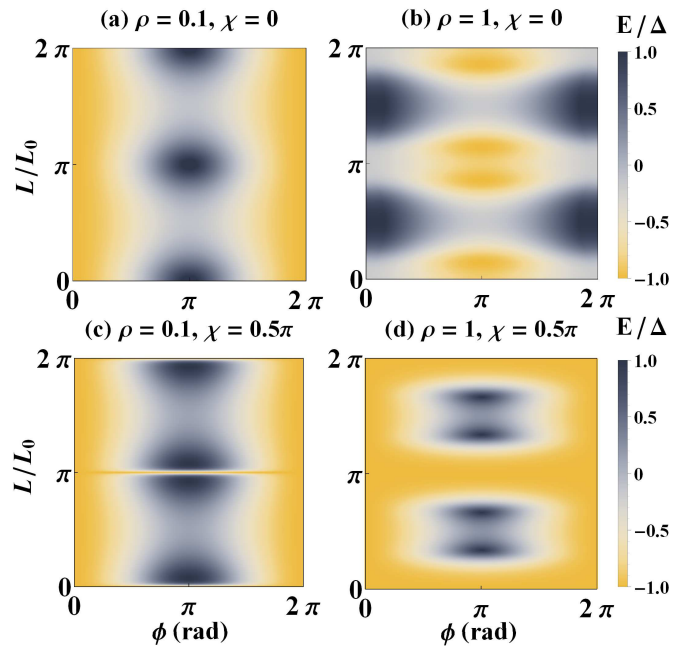}
\hspace{0.1cm}}
\caption{Density plot of variation of the ABS spectra with $\text{L}/\text{L}_0$ and $\phi$ for different choices of $\rho$ and $\chi$ considering $\beta = 2$ and $Z = 0.01$.}
\label{fig7}
\end{figure}

 Fig. \ref{fig7} displays the variation of gap energy as a function of $\text{L}/\text{L}_0$ and $\phi$ for spin mixing and spin-flipping processes. We observe a drastic change in the gap energy without spin flipping for weak and strong magnetic barriers. A signature of energy gap oscillation is observed with the change in $\text{L}/\text{L}_0$. Though the system retains the characteristics of gap oscillation of a conventional JJ for $\rho = 0.1$ but for $\rho  = 1$, the oscillation maxima are found for $(\phi, \text{L}/\text{L}_0)$ $\sim$ $(0, \frac{\pi}{2})$, $(0, \frac{3\pi}{2})$, $(2\pi, \frac{\pi}{2})$ and $(2\pi, \frac{3\pi}{2})$ as seen from  Figs. \ref{fig7}(a) and  Fig. \ref{fig7}(b). In the presence of spin flipping, similar characteristics are seen for $\rho = 0.1$ with $\chi = 0.5\pi$, but a sharp splitting is seen at $\text{L}/\text{L}_0 = \pi$. It is to be noted for a strong magnetic barrier with $\chi = 0.5\pi$, the oscillation maxima are observed at $\phi = \pi$, which suggests no change in the superconducting phase. Moreover, energy gap maxima shrinks and arise in the vicinity of $(\phi, \text{L}/\text{L}_0)$ $\sim$ $(\pi, \frac{\pi}{4})$, $(\pi, \frac{3\pi}{4})$, $(\pi, \frac{5\pi}{4})$ and $(\pi, \frac{7\pi}{4})$ as observed from Fig. \ref{fig7}(d).

The results of this work are also consistent with the results for the diffusive limit. It is primarily due to the following: (1) The $\phi$ states and thus the anomalous supercurrent originated due to symmetry of spin chirality and scattering from the spin active interfaces. Since these two effects are inherent, they are independent of ballistic and diffusive limits. (2) Suppression of higher harmonics resulting in the sinusoidal nature of CPR for low and high barrier transparency. So, our results are close to adequate in diffusive limit. One may use Matsubara Green's function methods \cite{pinon}, and quasi-classical Eilenberger \cite{miyawaki} equations to obtain the Andreev levels in diffusive limit. However, an analytic expression for ABS energies with all contributing parameters could be more complex and can be thus solved by numerical simulations. There may be some difference in the results in ballistic limits qualitatively. However, it is expected to have no drastic change overall.

\section{Conclusions and Proposed experimental realization }
In this work, we have extensively studied the quantum transport, formation of Andreev levels and Josephson supercurrent in an ISC$|$HM$|$ISC JJ. We have considered the effect of spin mixing, spin flipping and SOC of ISC on ABS spectra, CPR and tunnelling conductance in transparent and opaque barrier limits. Although a signature of splitting of the Andreev is noticed for spin mixing processes in the absence of SOC, the presence of SOC induces another splitting of the Andreev levels. Also, the $0 - \pi$ transition can be achieved at a low barrier magnetic moment. This is due to the spin chirality and lack of inversion symmetry of the ISC. Moreover, the effect of spin flipping is also observed in the presence of SOC. This results in a tunable $\phi$- phase JJ, which can be controlled via barrier magnetic moment, mismatch angle and SOC. Notably, the interplay of spin mixing and spin flipping processes with SOC can host Majorana modes in our proposed system. In summary, the fabrication problem of JJ's involving half metals has been reduced via SOC of the ISC. Another important aspect of this work is that $0 - \pi$ transition can be achieved even for an opaque barrier with suitable barrier moment and mismatch angle, which indicates the formation of possible $\phi$ junctions in diffusive limit also. A signature of splitting of the Andreev levels is observed for both single-band ($\mu_\text{S} < \beta$) as well as double-band ($\mu_\text{S} > \beta$). However, for $\mu_\text{S} = 0.5\beta$, the Andreev levels are degenerate. The unequal probabilities of Andreev reflected electrons and holes due to the presence of SOC give rise to finite sub-gap conductance. Moreover, the tunnelling conductance is found to be dependent on spin mixing and spin flipping processes. Furthermore, anomalous Andreev levels are observed for different HM length scales in the presence of spin mixing and spin flipping processes. 

Our work also focuses on the applications of our junction in quantum phase batteries, quantum computing, next-generation memories and other quantum devices. \cite{ioffe,ioffe2,hilgenkamp, madden}. The setup of our work, as shown in Fig. \ref{fig1}, can be easily realized in the laboratory. A similar structure without a spin-active barrier has already been reported \cite{cheng}. Another work has also studied a spin-active barrier structure via spin injection\cite{ouassou}. Recently, several experimental works indicate the possible fabrication of TMD-based ISC JJ \cite{hamil, ai, li,idzuchi,dvir,xiao2}. So, the fabrication of a spin-active ISC$|$HM$|$ISC JJ with TMDs as an ISC should not be difficult and can also be easily realized. Our work will readily help researchers experimentally design tunable phi junctions that can fabricate tunable phase batteries via spin injection without inter-facial magnetic inhomogeneity, enhancing the efficiency of next-generation quantum computing devices.
 
\appendix
\section{Matrix $\hat{\mathcal{A}}(E_\pm)$ for a spin-active ISC$|$HM$|$ISC Josephson junction}
The matrix $\hat{\mathcal{A}}(E_+)$ can be obtained by using the boundary conditions Eqs. (\ref{eq13})-(\ref{eq16}). An extensive form of the matrix $\hat{\mathcal{A}}$ for the ISC$|$HM$|$ISC Josephson junction can be written as
\begin{widetext}
\begin{multline}
\label{A1}
\hat{\mathcal{A}}(E_+)=\\\left(
\begin{array}{cccccccccccccccc}
 \Gamma _1 & 0 & 0 & 1 & -1 & 0 & 0 & 0 & -1 & 0 & 0 & 0 & 0 & 0 & 0 & 0 \\
 0 & \Gamma _1 & -1 & 0 & 0 & -1 & 0 & 0 & 0 & -1 & 0 & 0 & 0 & 0 & 0 & 0 \\
 0 & -1 & \Gamma _2 & 0 & 0 & 0 & -1 & 0 & 0 & 0 & -1 & 0 & 0 & 0 & 0 & 0 \\
 1 & 0 & 0 & \Gamma _2 & 0 & 0 & 0 & -1 & 0 & 0 & 0 & -1 & 0 & 0 & 0 & 0 \\
 0 & 0 & 0 & 0 & \delta _1 & 0 & 0 & 0 & \delta _3 & 0 & 0 & 0 & -\Gamma _{10} & 0 & 0 & -\Gamma _6 \\
 0 & 0 & 0 & 0 & 0 & \delta _2 & 0 & 0 & 0 & \delta _4 & 0 & 0 & 0 & -\Gamma _9 & \Gamma _5 & 0 \\
 0 & 0 & 0 & 0 & 0 & 0 & \delta _3 & 0 & 0 & 0 & \delta _1 & 0 & 0 & \Gamma _4 & -\Gamma _8 & 0 \\
 0 & 0 & 0 & 0 & 0 & 0 & 0 & \delta _4 & 0 & 0 & 0 & \delta _2 & -\Gamma _3 & 0 & 0 & -\Gamma _7 \\
 \kappa _+ \Gamma _1 & \Gamma _1 \Omega _2 & -\Omega _2 & \kappa _- & -i \kappa _+ & 0 & 0 & 0 & i \kappa _+ & 0 & 0 & 0 & 0 & 0 & 0 & 0 \\
 \Gamma _1 \Omega _2 & \Gamma _1 \mathcal{Q}_{11} & -\mathcal{Q}_{12} & \Omega _2 & 0 & -i \kappa _- & 0 & 0 & 0 & i \kappa _- & 0 & 0 & 0 & 0 & 0 & 0 \\
 \Omega _2 & -\mathcal{Q}_7 & \Gamma _2 \mathcal{Q}_8 & \Gamma _2 \Omega _2 & 0 & 0 & i \kappa _+ & 0 & 0 & 0 & -i \kappa _+ & 0 & 0 & 0 & 0 & 0 \\
 \mathcal{Q}_9 & -\Omega _2 & \Gamma _2 \Omega _2 & \Gamma _2 \mathcal{Q}_{10} & 0 & 0 & 0 & i \kappa _- & 0 & 0 & 0 & -i \kappa _- & 0 & 0 & 0 & 0 \\
 0 & 0 & 0 & 0 & \delta _1 \mathcal{Q}_1 & \delta _2 \Omega _2 & 0 & 0 & \delta _3 \mathcal{Q}_3 & \delta _4 \Omega _2 & 0 & 0 & -i k_+ \Gamma _{10} & 0 & 0 & i k_+ \Gamma _6 \\
 0 & 0 & 0 & 0 & \delta _1 \Omega _2 & \delta _2 \mathcal{Q}_2 & 0 & 0 & \delta _3 \Omega _2 & \delta _4 \mathcal{Q}_4 & 0 & 0 & 0 & -i k_- \Gamma _9 & -i k_- \Gamma _5 & 0 \\
 0 & 0 & 0 & 0 & 0 & 0 & \delta _3 \mathcal{Q}_3 & \delta _4 \Omega _2 & 0 & 0 & \delta _1 \mathcal{Q}_1 & \delta _2 \Omega _2 & 0 & i k_- \Gamma _4 & i k_- \Gamma _8 & 0 \\
 0 & 0 & 0 & 0 & 0 & 0 & \delta _3 \Omega _2 & \delta _4 \mathcal{Q}_4 & 0 & 0 & \delta _1 \Omega _2 & \delta _2 \mathcal{Q}_2 & -i k_+ \Gamma _3 & 0 & 0 & i k_+ \Gamma _7 \\
\end{array}
\right)
\end{multline}
\end{widetext}
where we define,\\
$\Gamma_1=w e^{i \phi_\text{L}}$, \\
$\Gamma_2=w e^{-i \phi _\text{L}}$,\\
$\Gamma_3=e^{i k_+ \text{L}}$, \\
$\Gamma_4=e^{i k_- \text{L}}$,\\
$\Gamma_5=e^{-i k_- \text{L}}$, \\
$\Gamma_6=e^{-i k_+ \text{L}}$,\\
$\Gamma_7=w e^{-i \left(k_+ \text{L}+\phi _\text{R}\right)}$,\\ 
$\Gamma_8=w e^{-i \left(k_- \text{L}+\phi _\text{R}\right)}$,\\
$\Gamma_9=w e^{i \left(k_- \text{L}+\phi _\text{R}\right)}$, \\
$\Gamma_{10}=w e^{i \left(k_+ \text{L}+\phi _\text{R}\right)}$,
\\\\
$\delta_1=e^{i \kappa _+ \text{L}}$, \indent\indent $\delta _2=e^{i \kappa _- \text{L}}$,\\
$\delta_3=e^{-i \kappa _+ \text{L}}$, \indent\hspace{0.5mm} $\delta _4=e^{-i \kappa _- \text{L}}$,\\\\
$\mathcal{Q}_1= Z+\Omega _1+i \kappa _+$, \indent\indent $\mathcal{Q}_2=Z-\Omega _1+i \kappa _-$ \\
$\mathcal{Q}_3= Z+\Omega _1-i \kappa _+$, \indent\indent $\mathcal{Q}_4=Z-\Omega _1-i \kappa _-$ \\
$\mathcal{Q}_5= Z+\Omega _1-i k_+$, \indent\indent $\mathcal{Q}_6=Z+\Omega _1+i k_+$ \\
$\mathcal{Q}_7= Z+\Omega _1-i k_-$, \indent\indent $\mathcal{Q}_8=Z+\Omega _1+i k_-$ \\
$\mathcal{Q}_9= Z-\Omega _1-i k_+$, \indent\indent $\mathcal{Q}_{10}=Z-\Omega _1+i k_+$ \\
$\mathcal{Q}_{11}= Z-\Omega _1-i k_-$, \,\,\ \indent $\mathcal{Q}_{12}=Z-\Omega _1+i k_-$ \\

\end{document}